\documentclass[twocolumn ,showpacs,preprintnumbers,amsmath,amssymb,prl, superscriptaddress]{revtex4}

\usepackage{graphicx}
\usepackage{dcolumn}
\usepackage{bm}

\usepackage{color,ulem}

\begin{document}
\title{Ultrafast dephasing of light in strongly scattering GaP nanowires.}
\author{Martina Abb}
\affiliation{SEPnet and the Department of Physics and Astronomy, University of
Southampton, Highfield, Southampton SO17, 1BJ, United
Kingdom}\email{O.Muskens@soton.ac.uk}
\author{Erik P. A. M. Bakkers}
\affiliation{Eindhoven University of Technology, P.O. Box 513, 5600 MB Eindhoven, The Netherlands}
\author{Otto L. Muskens}
\affiliation{SEPnet and the Department of Physics and Astronomy, University of
Southampton, Highfield, Southampton SO17, 1BJ, United
Kingdom}\email{O.Muskens@soton.ac.uk}

\date{\today}

\begin{abstract}
We demonstrate ultrafast dephasing in the random transport of
light through a layer consisting of strongly scattering GaP
nanowires. Dephasing results in a nonlinear
intensity modulation of individual pseudomodes which is 100 times larger than
that of bulk GaP. Different contributions to the nonlinear
response are separated using total transmission, white-light
frequency correlation, and statistical pseudomode analysis. A
dephasing time of $1.2\pm 0.2$~ps is found. Quantitative agreement
is obtained with numerical model calculations which include
photoinduced absorption and deformation of individual scatterers.
Nonlinear dephasing of photonic eigenmodes opens up avenues for ultrafast
control of random lasers, nanophotonic switches, and photon
localization.
\end{abstract}
\maketitle

The appearance of an interference pattern after transport of
coherent light through a multiple scattering medium is the result
of coherent summation of thousands of light paths with random
phases. In recent years, new methods have been developed to exploit
the coherent aspects of diffuse light transport for imaging
\cite{Fink2007,Vellekoop2010}. Classical waves, like light or
sound, provide unique opportunities to study mesoscopic wave
transport, such as localization, in the absence of many-body
interactions usually found for electrons \cite{Lagendijk2009,
Hu2008}. Coherent light scattering has recently lead to exciting
new directions of research, such as coupling of localized states
with quantum emitters \cite{Sapienza2010}, transverse localization,
and combined effects of localization and nonlinearity
\cite{Schwarz2000}.

Although the lack of many-body interactions makes phase coherence
generally more robust for optical waves than for electrons,
several dephasing processes can influence the transport of light,
such as magnetic fields, scattering from moving particles, and
polarization effects
\cite{Golubentsev,AkkermansBook,ErbacherEPL93}. The time required
for electrons or photons to loose their coherence due to inelastic
collisions is known in the theory of electronic conductance as the
phase breaking time \cite{Altshuler81}. While for electrons phase
breaking (decoherence) processes have only been measured
indirectly through changes in the conductivity with temperature,
dephasing can be observed directly for optical waves by the
changes in speckle pattern \cite{Kaveh87}, which is used in
diffusing wave spectroscopy \cite{Pine88}.

In this Letter, we demonstrate a new regime of reversible
dephasing in a random medium on ultrafast time scales. The
ultrafast dephasing takes place a million times faster than other
dephasing mechanisms such as Brownian movement, opening up new
applications in ultrafast control of random lasers, quantum
emitters, and localization phenomena. The material under study
consists of a layer of semiconductor nanowires, which are
important for novel applications in optoelectronics and
photovoltaics \cite{Yang,Lieber,Atwater}. We have recently
demonstrated that these semiconductor nanowires form one of the
most efficient light trapping layers available to date
\cite{Muskens2009}. Here, we combine these favorable scattering
properties with the intrinsic nonlinearity of the semiconductor
host material to achieve active control on ultrafast time scales.

Gallium phosphide (GaP) nanowires were grown using
metallo-organic vapour phase epitaxy as described in
Ref.~\cite{Muskens2009}. The length of the nanowires was
controlled by the VLS growth time, while the thickness and volume
fraction was independently tuned by switching to a lateral growth
regime where material was deposited homogeneously over the
nanowire sidewalls. In the experiments of this work, we make use
of two nanowire samples, respectively with mean free path $\ell=0.16\pm 0.02$~$\mu$m and thickness $L=1.5$~$\mu$m (sample 10 of
Ref.~\cite{Muskens2009}, here sample 1), and $L=4.0$~$\mu$m, $\ell=0.2\pm 0.02$~$\mu$m (sample 11 of
Ref.~\cite{Muskens2009}, here sample 2), at a wavelength of 632~nm. The variation of $\ell$ with wavelength was found to be small over the spectral range of our experiments \cite{Muskens2009}.

\begin{figure}[t]
\centering
\includegraphics[width=7.6cm]{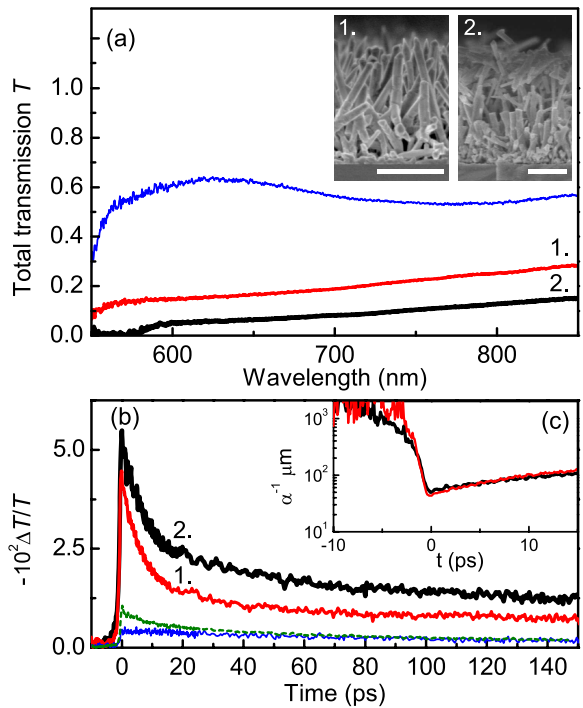}
\caption{\label{fig:totaltransmission} (color online) (a) Total
transmission spectra for nanowire layers 1 and 2, and for a GaP substrate (thin line, blue). Inset: scanning electron microscopy images, scale
bars 1~$\mu$m. (b) Pump-probe total transmission $\Delta T/T$ for
the samples of (a) and for a 51-nm diameter nanowire layer (dash, sample 6 of Ref.~\cite{Muskens2009}). (c) Absorption length $\alpha^{-1}$ obtained from (b) using Eq.~(\ref{eq:ohm}).}
\end{figure}

We investigated the photoinduced nonlinear response using a broadband white-light supercontinuum probe
generated by focusing of 800-nm laser pulses from a regenerative
amplifier (Coherent RegA, 250 kHz, $200\pm 10$~fs pulse duration)
in a quartz window. The material is pumped using the second harmonic at 400 nm wavelength. Both the pump and
probe are focused onto the sample using a lens with a 6-cm focal
length, resulting in a focus of around 5 $\mu$m in diameter and a pump fluence of 8~mJ/cm$^2$. Total transmission measurements are shown in
Fig.~\ref{fig:totaltransmission}(a) for the nanowires as well as
for a GaP substrate (thin line, blue).

In absence of
absorption, the transmission through a scattering layer follows
Ohm's law $T\simeq \ell/L$. The total transmission can be derived
including effects of internal reflection \cite{GomezRivas}, which
for small absorption ($L_a>L>\ell$) reads
\begin{equation} \label{eq:ohm}
T= \frac{\ell+z_{e1}}{(1+z_{e1}z_{e2}/L_a^2)L+(z_{e1}+z_{e2})} \,
,
\end{equation}
where $L$ denotes the slab thickness, $z_{e,i}$ are extrapolation
lengths which represent the effects of internal reflection at the
front and rear interfaces of the diffuse slab. The diffuse
absorption length is defined as $L_a=(\ell /3\alpha)^{1/2}$, with
$\alpha^{-1}$ the bulk absorption length.

Pump-probe total transmission measurements are shown in
Fig.~\ref{fig:totaltransmission}(b). All samples show a reduction
of the total transmission directly following excitation by the
pump laser. Independent experiments were performed verifying that
mesoscopic contributions to the total transmission, i.e. fluctuations caused by a finite number of transmission channels \cite{Scheffold98}, were unimportant under the present experimental conditions. Therefore the signals in Fig.~\ref{fig:totaltransmission}(b) are attributed
to changes in the absorption coefficient of the materials. Transient absorption in semiconductor nanostructures results from free carrier generation \cite{ElSayed}. A
faster relaxation time is observed for nanowires
compared to the bulk semiconductor, which can be attributed to
trapping of photogenerated carriers at surface states \cite{Hartland2009}. The transient relative absorption increases from $3\times 10^{-3}$ for the GaP substrate to $5.2\times 10^{-2}$ for nanowire sample 2. We attribute this strong increase in absorption by light trapping in the nanowire layer \cite{Muskens2008}. Light trapping can be assessed from the time the light resides in the micrometer-thin photoexcited region, where it is susceptible to nonlinear absorption. For the GaP substrate, the coherent beam traverses this region in around $10$~fs, while, for nanowire sample 2, the light diffusion time amounts to approximately $L^2/\pi^2 D\simeq 0.15$~ps. We note that optically thin nanowires [dashed line, green in Fig.~\ref{fig:totaltransmission}(b)] did not show such a marked increase of the nonlinear absorption, eliminating other possible nanowire surface effects.

 Equation \ref{eq:ohm} can be inverted to yield values for the absorption length $\alpha^{-1}$ for a pump-probe signal $\Delta T/T$, as shown in Fig.~\ref{fig:totaltransmission}(c). A minimum absorption length  $\alpha^{-1}$ of $49.9 \pm
4$~$\mu$m is found, corresponding to a change of the imaginary
refractive index $\Delta \kappa = c_0\alpha/4 \pi \nu$ of $(1.0
\pm 0.2) \times 10^{-3}$. Using a free-carrier
absorption cross section of GaP of $1.1\times 10^{-18}$~cm$^{-2}$
\cite{Rychnovsky94}, our value for $\Delta \kappa$ amounts to a
free-carrier density of $2\times 10^{20}$~cm$^{-3}$, which is a
realistic density for the strong optical pumping regime under
study.

\begin{figure}[t]
\centering
\includegraphics[width=7.8cm]{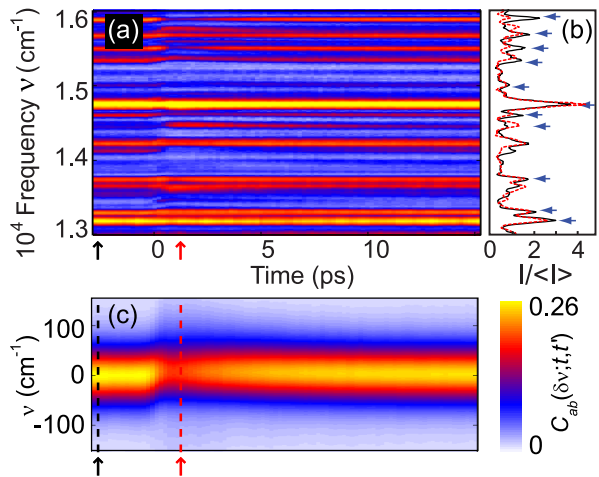}
\caption{\label{fig:c1} (color online) (a) Map of time-resolved
frequency speckle for sample 2, with (b) spectra taken at $-3$~ps
(black) and $1$~ps (red) pump-probe time delay. Blue arrows
indicate peaks selected for pseudomode analysis. (c) Map of
experimental correlation $C_{ab}(\nu;t,t')$.}
\end{figure}

In the following we focus on the results for nanowire sample 2.
Similar results are obtained for sample 1 \cite{suppl}. To assess the effects of nonlinear pumping on the
dynamical transport parameters of the random medium, we measured
the frequency correlations of light transmitted through the
nanowire slab. For this purpose, a cone of light of around
$2^\circ$ angular width was collected and analyzed using a fiber
spectrometer. Broadband linear polarization filters
were used both for the incident and for the detected light.
Spectra were measured as a function of pump-probe delay time,
yielding time-resolved spectral maps such as shown in
Fig.~\ref{fig:c1}(a). The transmission spectra consist of
large fluctuations corresponding to frequency speckle. Within 1~ps
following excitation with a pump laser, changes occur in both the
spectral position and amplitude of the fluctuations, as is
illustrated in Fig.~\ref{fig:c1}(b) where we compare two
spectra taken respectively 3~ps before and 1~ps after arrival of
the pump pulse. To quantify these variations in the speckle pattern, we calculated the time-correlation function $C_{ab}(\nu;t,t')$ from the
time-resolved spectra, using the conventional definition of the intensity-intensity correlation \cite{Feng94}. The spectral cross-correlation was
calculated for each spectrum at time $t$ with the spectrum at $t'=-3$~ps, as shown in Fig.~\ref{fig:c1}(c), where spectra taken at 20 different locations on the sample were used to obtain an ensemble average.

\begin{figure}[t]
\centering
\includegraphics[width=7.5cm]{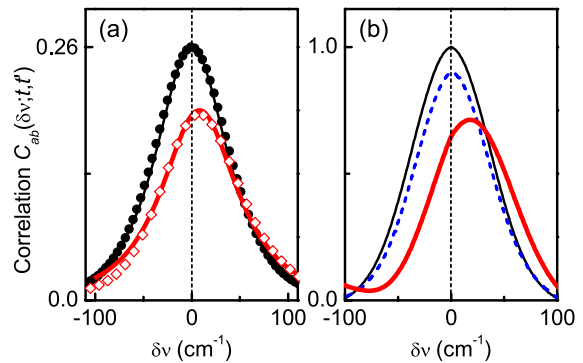}
\caption{\label{fig:c1traces} (color online) (a) Cross sections of
Fig.~\ref{fig:c1}(c) at delay times of $-3$~ps (dots, black),and
$1$~ps (diamonds, red). Lines represent fits using
Eq.~(\ref{eq:c1model}), without (black) and with (red line)
dephasing model with $\tau_\gamma=1.2 \pm 0.2$~ps. (b) Finite-element
calculations \cite{suppl} of autocorrelation (line, black) and
cross-correlations for a refractive index change $\Delta n_{\rm
GaP}=-2\times 10^{-3}+10^{-3} i$ (dashed line, blue), and an
additional $10^{-3}$ size increase of individual scatterers (thick
line, red).}
\end{figure}

For times directly after the arrival of
the pump pulse, the time-correlation shows a reduction in amplitude, accompanied
by a spectral shift, denoted hereafter by $\delta \nu_0$, of $10.2
\pm 1.0$ cm$^{-1}$. Cross sections of Fig.~\ref{fig:c1}(c) are shown in Fig.~\ref{fig:c1traces}(a) at -3~ps (black dots) and $1$~ps
(red diamonds). The spectral autocorrelation can be
calculated using \cite{Feng94}
\begin{equation} \label{eq:c1model}
C_{ab}(\Delta \nu, \tau_a)\simeq N[|\eta
L|^2/|\sinh(\eta L)|^2] \, ,
\end{equation}
with $\eta=D^{-\frac{1}{2}}(\tau_a^{-1} + 2\pi i \Delta\nu)^{1/2}$, where $D$ is the diffusion constant of the
light, and $\tau_a=(v_{E} \alpha)^{-1}$ the absorption time,
$v_E=3D/\ell$ being the energy velocity. The
normalization constant $N$ is used to fit Eq.~(\ref{eq:c1model}) to
our data given the reduced experimental speckle contrast.
Equation~(\ref{eq:c1model}) gives good agreement
for the autocorrelation at -3~ps [line, black in
Fig.~\ref{fig:c1traces}(a)] for $D=15 \pm 2$~m$^2$/s.

The reduction of the correlation between spectra at different
times allows for an estimate of the dephasing
time. As shown in Ref.~\onlinecite{AkkermansBook},
dephasing processes affect the average
probability of light transport through an exponential decay
$\exp(-t/\tau_\gamma)$ of the correlation between two amplitudes,
where $\tau_\gamma$ is the dephasing time. The effects of
dephasing and average refractive index change
\cite{Faez09} can thus be introduced into the correlation
function by replacing $\eta$ in Eq.~(\ref{eq:c1model}) by the new
\begin{equation}
\label{eq:etagamma}
\eta_\gamma = D^{-\frac{1}{2}}\left[ \tau_a^{-1}
+ \tau_\gamma^{-1} + 2\pi i(\Delta\nu+\nu \frac{\Delta n_{e}}{n_{e}})\right]^{1/2}.
\end{equation}
We note that the normalization factor $N$ only depends on the variance of the two individual configurations, which is not affected by the dephasing or the effective index change. We fitted
the reduction in correlation amplitude by the dephasing model of Eq.~(\ref{eq:etagamma}). Good
quantitative agreement is found [thick line, red in
Fig.~\ref{fig:c1traces}(a)] for a dephasing time $\tau_\gamma=1.2
\pm 0.2$~ps and a change in the average refractive index of the
scattering medium of $\Delta n_{e}/n_{e}=(-5.8\pm 2)\times
10^{-4}$.

\begin{figure}[t]
\centering
\includegraphics[width=8.7cm]{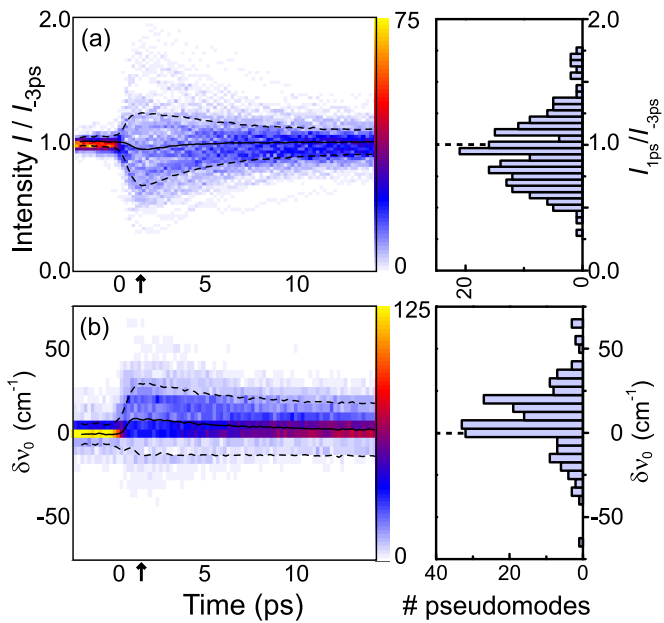}
\caption{\label{fig:pseudomodestat} (color online) Results of
pseudomode analysis of 202 modes, yielding nonlinear intensity
change $I/I_{\rm -3ps}$ (a) and spectral shift $\delta \nu_0$ (b).
(Red dashed lines) statistical average of the distributions.
Arrows indicate time location of the histograms shown in
right-hand panels. Lines indicate averages
(solid) and standard deviations (dash) of the distributions.}
\end{figure}

We considered, next to the loss of correlation, also the effect of ultrafast excitation on individual pseudomodes. Pseudomodes are defined as the transmission modes of the system, which take on the form of spectral resonances, with a spectral width determined by the leakage of the state to free
space. The spectrum of pseudomodes plays an important role in theories of
localization \cite{Thouless,SkipetrovPRL06} and random lasers \cite{Ling2001}. We have traced the time evolution of a subset of pseudomodes selected from our experimental spectra using an algorithm \cite{remark2}, including modes such as those indicated by the blue arrows in Fig.~\ref{fig:c1}(b). The position and amplitude of each pseudomode was tracked as function of the pump-probe delay time, yielding a set of well-defined intensity changes and spectral shifts shown in Fig.~\ref{fig:pseudomodestat}(a,b). Each time delay $t$ represents a
histogram, such as is shown in the right-hand panels of
Fig.~\ref{fig:pseudomodestat}(a,b) for $t=1$~ps. A wide distribution of spectral and intensity variations is found with both decreasing and increasing values, which extend beyond the variation of the time-correlation in Fig.~\ref{fig:c1}(c). A fraction of modes in Fig.~\ref{fig:pseudomodestat}(b) shows a negative frequency shift, which cannot be explained by an average refractive index change but which indicates a more chaotic change of the pseudomode spectrum.


We have investigated possible dephasing processes using a
numerical finite-element (COMSOL) model of light propagation in a two-dimensional
slab consisting of a random arrangement of dielectric cylinders
\cite{suppl}. The frequency cross-correlation was calculated between two
configurations corresponding to the situations before and after excitation. Consistent with earlier
observations \cite{Faez09}, no decorrelation was
found for only a small refractive change of around $10^{-3}$.
However, our model calculations identified two other possible
dephasing mechanisms: absorption and particle deformation. While
absorption [dashed line, green in Fig.~\ref{fig:c1traces}(b) for
$\Delta n_{\rm GaP} = -2\times10^{-3} + 10^{-3} i$] does not change
the phase of individual light paths, it does lead to a
redistribution of phasor weights in the total speckle field by
suppression of contributions from long paths.
This interpretation is only partially correct in
time-resolved experiments, where paths are separated in time and only the finite duration of the probe pulse
leads to mixing of paths with different amounts of absorption. We point out that inhomogeneous absorption may also lead to variations
between different paths of equal length and thus to dephasing.

We explored another possible dephasing mechanism
caused by stress-induced deformations. It was found that a
$10^{-3}$ increase of the scatterer size is sufficient to achieve
quantitative agreement with experiment [thick line, red in
Fig.~\ref{fig:c1traces}(b)]. Such deformations are likely to occur
under the typical conditions of high power pulsed laser excitation
of solids \cite{Maris}, and may be enhanced by the particular
morphology of the nanowires \cite{Mariager10}.
The effect of deformations is important as it
represents the contribution of changes in the scattering matrix of
individual scatterers. No dephasing was observed
when randomly displacing each scatterer over 0.1~nm.

Our analysis of pseudomode distributions and spectral correlations
reveals that ultrafast dephasing enables reversible switching
between random configurations involving thousands of scattering
events. Although the intensity and frequency changes appear
chaotic, they are highly reproducible over millions of pump-probe
cycles. The observed pseudomode intensity
modulations of up to 50\% are large for the field of
nanophotonics, and comparable to e.g. nonlinear switching of
three-dimensional photonic crystals \cite{Euser05}. The similar decorrelation values for the two nanowire layers with different thicknesses \cite{suppl} indicates that dephasing is limited by the pump absorption length, rather than by sample thickness; further
optimization of dephasing using e.g. multiphoton excitation \cite{Euser05}, will
be of importance for exploitation of these effects in
applications. Deterministic control of pseudomodes may be combined
with active gain media to produce novel types of ultrafast random
lasers where the lasing modes can be manipulated using an external
control pulse. The dephasing time of $\sim 1.2$~ps is of the order
of the transport time through the layer, which paves the way for
nonadiabatical control over light paths. Of particular interest
here will be the possibility to break the phase coherence of
correlated paths involved in mesoscopic transport, and ultimately
of time-reversed light paths involved in photon localization.

We thank the ORC FASTlab and D. Kundys for support, and A.
Lagendijk and S. Faez for fruitful discussions. This work was supported by
EPSRC through grant EP/H019669/1.



\begin{thebibliography}{10}
\bibitem{remark2} Modes were selected using an algorithm based on the condition that the normalized intensity shows a peak larger than 0.5 above its surroundings over a spectral range equal to two times the correlation width.
\end{thebibliography}
\end{document}